\documentstyle[epsfig,12pt]{article}

\newcommand{\fig}[1]{Fig.\ref{#1}}
\newcommand{\tab}[1]{table \ref{#1}}

\newcommand{\eqn}[1]{Eq.(\ref{#1})}
\def\gev{\mathrm{GeV}}

\newcommand{\ben}{\begin{enumerate}}
\newcommand{\een}{\end{enumerate}}
\newcommand{\bit}{\begin{itemize}}
\newcommand{\eit}{\end{itemize}}
\newcommand{\bc}{\begin{center}}
\newcommand{\ec}{\end{center}}
\newcommand{\bb}{\begin{bf}}
\newcommand{\eb}{\end{bf}}
\newcommand{\bsm}{\begin{small}}
\newcommand{\esm}{\end{small}}
\newcommand{\bns}{\begin{normalsize}}
\newcommand{\ens}{\end{normalsize}}
\newcommand{\bq}{\begin{equation}}
\newcommand{\eq}{\end{equation}}
\newcommand{\bqa}{\begin{eqnarray}}
\newcommand{\eqa}{\end{eqnarray}}

\newcommand{\vb}{\vspace*{2cm}}

\newcommand{\nn}{\nonumber}

\def\cgp{C.G.~Pa\-pa\-do\-pou\-los}

\def\ena{E.N.~Ar\-gy\-res}

\def\tg{\mbox{tg~}}
\def\ctg{\mbox{ctg~}}
\def\e{\varepsilon}
\def\c2{\chi^2}
\def\SM{Standard Model\ }

\def\L{{\cal L}}
\def\O{{\cal O}}
\def\onehalf{{1\over 2}}
\def\vtau{\mbox{\boldmath $\tau$}}
\def\vW{\mbox{\boldmath $W$}}
\def\kg{\kappa_\gamma}\def\kZ{\kappa_Z}
\def\xg{x_\gamma}\def\xZ{x_Z}
\def\dZ{\delta_Z}\def\lg{\lambda_\gamma}
\def\lZ{\lambda_Z}

\begin{document}

\pagestyle{empty}

\begin{flushright}CERN-TH.7157/94\\
\end{flushright}

\vspace*{2cm}
\bc\begin{Large}
{\bf Single-$W$ versus $W$-pair production at
LEP~II}\\ \end{Large}
\vspace*{1cm}
{\large Costas G.~Papadopoulos \\
CERN-Theory Division, Geneva, Switzerland} \\[24pt]
\vb
ABSTRACT\\[12pt]  \ec
\begin{quote}
This paper is a study of the single-$W$ and $W$-pair
production at LEP~II.
I show that the sensitivity of the single-$W$ production
channel on the anomalous trilinear couplings of $W$ is essentially
the same as the sensitivity of the channel $e^+e^-\to W^+W^-$.
Therefore, even without the reconstruction of the $W$ boson,
deviations form the \SM predictions on
trilinear couplings can be measured at the level of $0.2 - 1$,
using single-$W$ production at LEP~II.
\end{quote}
\vfill
\begin{flushleft} CERN-TH.7157/94\\
February 1994
\end{flushleft}
\newpage
\pagestyle{plain}
\setcounter{page}1
\par Electroweak precision tests at the $Z^0$ pole using LEP~I
data, constrain in a rather important way the
physics that can be probed in the near future \cite{alta,rola}.
Taking into account also
the Tevatron results \cite{teva}, it will be very difficult
to study directly the top quark
or the Higgs scalar particle at LEP~II. If this is true, then
LEP~II experiments should focus on $W$-physics.
Moreover, given the limiting rates expected at LEP~II energies
(in contrast with LEP~I), it will be rather
difficult to have a new precision-test experiment.
Therefore the main, well-defined area of experimental
and phenomenological research will be the study
of the interactions of $W$, and especially the
measurement of the trilinear and quadrilinear couplings
among the gauge bosons \cite{gouna:1,hagi:lep2,lep2,papa:sngw,papa:lep2}.
\par
The interest of these measurements lies in the fact
that the \SM prediction for them is strongly related
to the existence of the underlying symmetry: its non-Abelian
character will be tested directly by measuring these couplings.
On the other hand any `new physics' manifests itself at low energies
by modifying these couplings: the difference between different
models is the strength of this modification. In particular large
deviations from \SM predictions will point in
the direction of
a composite structure, whereas small ones can be accommodated by
renormalizable theories.
\par
In order to study the anomalous couplings one usually
introduces an interaction Lagrangian \cite{hagi:lep2},
which in the unitary (physical) gauge and
considering only CP-conserving couplings,
is written in the following form:
\bqa
\L&=&\sum_{V=\gamma,Z}\biggl\{g_V(
V_\mu W^{-\mu\nu}W^{+\nu}
-V_\mu W^{+\mu\nu}W^{-\nu}
+\kappa_V V_{\mu\nu}W^{+\mu}W^{-\mu\nu})\nn\\
&+&g_V{\lambda_V\over M_W^2}V_{\mu\rho}W^{+\rho\nu}W^{-\mu}_\rho
\biggr\}\eqa
where
\bqa W^+_{\mu\nu}=\partial_\mu W^+_{\nu}-\partial_\nu W^+_{\mu}
\nn\eqa
$W^{\pm}$ is the $W$-boson field, and
$g_\gamma=e$, $g_Z=e \ctg\theta_w$, $\kg=\kZ=1$ and
$\lg=\lZ=0$ at tree order in the \SM.
It is more convenient to express the different couplings
in terms of their deviations form the \SM values. For this we
define the following deviation parameters:
\bqa
\dZ&=&g_Z-\ctg\theta_w\nn\\
\xg&=&\kg-1\nn\\
\xZ&=&(\kZ-1)(\ctg\theta_w+\dZ)
\label{anom}
\eqa
It is worth while to note that the interaction Lagrangian
becomes linear with respect to the above parameters (including also
$\lg$ and $\lZ$).
\par
During the last years, considerable progress has been achieved
concerning the understanding of the physics underlying
the anomalous couplings.
As Gounaris and Renard \cite{gouna:10} showed,
these anomalous couplings can be parametrized in a
manifestly gauge-invariant (but still non-renormalizable)
way: gauge-invariant operators involving higher-dimensional
interactions among gauge bosons and Higgs field are hidden
beyond them. It is the dynamics of the unknown Higgs sector
that is essentially probed by studying the trilinear couplings
of the massive vector bosons.
Precise electroweak tests at LEP~I already
impose severe restrictions on the possible operators
considered \cite{ruju,burg}.
Operators contributing to the $W$ to $Z$
mass ratio, and thus affecting the so-called $\rho$ parameter,
which has been measured with very high accuracy at LEP~I, should
be excluded.
Moreover, these gauge-invariant operators are multiplied by
powers of $v^2/\Lambda^2$, where $v=(\sqrt{2G_F})^{-1/2}$, is the
vacuum expectation value of the Higgs field and $\Lambda$
the scale where new physics is expected to appear.
In the limit $v/\Lambda\to 1$, operators with arbitrary dimension
have to be considered:
this is a `democratic' scenario, which in the language
of the old parametrization means that all anomalous
couplings are treated as independent. Of course, this scenario
suffers several drawbacks. First of all
it will be difficult experimentally to disentangle the
different contributions. At the same time, from a theoretical
point of view, it seems to be rather unexpected,
the new physics scale to be of the order of the
electroweak scale\footnote{However,
it is fair to emphasize that at higher energies
\cite{1loop:papa,bmt:500},
when it could be possible to test the
one-loop corrections in the \SM, a `democratic' scenario
has to be considered.}.
On the other hand, in the limit $v/\Lambda << 1$,
one could have a `hierarchical' scenario \cite{boud:1}, which
considerably simplifies
both experimental and theoretical
studies of anomalous couplings.
Adopting this last scenario, and restricting ourselves to
$SU(2)_L\times U(1)_Y$-invariant operators with dimension up to 8,
we can have the following list:
\bqa
\O_{B\Phi}&=&B^{\mu\nu}(D_\mu\Phi)^\dagger(D_\nu\Phi)\nn\\
\O_{W\Phi}&=&(D_\mu\Phi)^\dagger\;\vtau\cdot\vW^{\mu\nu}
(D_\nu\Phi)\nn\\
\O_W&=&{1\over 3!}(\vW^{\mu}_{\;\rho}\times\vW^{\rho}_{\;\nu})
\cdot\vW^{\nu}_{\;\mu}\nn\\
\O^\prime_{W\Phi}&=&(\Phi^\dagger\vtau\cdot\vW^{\mu\nu}\Phi)
(D_\mu\Phi)^\dagger(D_\nu\Phi)\nn\\
\O^\prime_W&=&(\vW^{\mu}_{\;\rho}\times\vW^{\rho}_{\;\nu})\cdot
(\Phi^\dagger{\vtau\over 2}\Phi)B^{\nu}_{\;\mu}
\label{oper}
\eqa
where $\tau_i=\onehalf\sigma_i$ (the Pauli matrices),
\bqa
B_{\mu\nu}=\partial_\mu B_{\nu}-\partial_\nu B_{\mu}
\nn\eqa
where $B_\mu$ is the $U(1)_Y$ gauge field,
\bqa
\vW_{\mu\nu}=\partial_\mu \vW_{\nu}-\partial_\nu \vW_{\mu}
-g_2\vW_{\mu}\times\vW_{\nu}\nn\eqa
where $\vW$ are the $SU(2)_L$ gauge fields and
\bqa
\Phi=\left( \begin{array}{c} \phi^+\\
{1\over \sqrt{2}}(v+H+i\phi^0)\end{array}\right)
\nn\eqa
is the Higgs doublet.
The covariant derivative $D_\mu$ is given, as usual, by
\bqa
D_\mu=\partial_\mu+i\; g_2\vtau\cdot\vW_\mu-i\; g_1 B_\mu
\nn\eqa
and $e=g_2\sin\theta_w=g_1\cos\theta_w$.
\par
The interaction Lagrangian can be written now as
\bqa
\L=\L_{SM}+\sum g_i\O_i
\eqa
where $\O_i$ are the operators given in \eqn{oper}.
The different contributions of the above operators to the
anomalous couplings \eqn{anom} are summarized in \tab{tab1}.
In the present analysis I only consider one $g_i$ coupling
at a time
to be non-zero. Since at LEP~II energies
it is not expected to test the one loop corrections
to the trilinear couplings, this approach is justified.
At higher energies,
where these corrections could be tested, one should consider
the case that all $\O_i$ are equally, or almost equally,
contributing to the effective Lagrangian.
This is because the radiative corrections to the trilinear couplings
cannot be parametrized in terms of a single operator: for instance
one-loop relations between $\kg$ and $\kZ$ are
energy dependent \cite{1loop:papa} and are generically
different from the relations given in \tab{tab1}.
\begin{table}[htb]
\bc\begin{tabular}{|c|c|}
\hline
$\O_{W\Phi}$ & $\xg\neq 0$, $\dZ={\xg\over s_w c_w}$,
$x_Z=-\xg\tg\theta_w$\\ \hline
$\O_{B\Phi}$ & $\xg\neq 0$, $x_Z=-\xg\tg\theta_w$\\ \hline
$\O_{W}$ & $\lg\neq 0$, $\lZ=\lg$\\ \hline
$\O^\prime_{W\Phi}$ & $\xg\neq 0$, $\xZ=\xg\ctg\theta_w$\\ \hline
$\O^\prime_{W}$ & $\lg\neq 0$, $\lZ=\lg\mbox{tg}^2\theta_w$\\ \hline
\end{tabular}
\caption[.]{Contribution of different gauge-invariant operators
to the anomalous couplings
($s_w=\sin\theta_w$ and $c_w=\cos\theta_w$).}
\label{tab1}
\ec\end{table}
\par In this paper I study the single-$W$ and $W$-pair
production and their sensitivity with respect to the
anomalous trilinear couplings as defined above.
By single-$W$ production I mean the process
$e^+e^-\to W^{\pm}\ell^{\mp}\nu_\ell$, where $\ell$ can be
either an electron or a muon.
In the case of a muon, it is easy to see that the main
contribution is expected to come from the region
where the muon and the corresponding neutrino
have a mass very close to the mass of the $W$: the so-called
resonant graphs are dominant.
On the other hand,
the electron case is expected to be more interesting
since $t$-channel graphs are also present and therefore
the non-resonant contribution should be important.
In the rest of the analysis I focus on the electron channel,
using the angular distribution of the electron as the
main tool to investigate the sensitivity of the single-$W$
production with respect to the anomalous couplings.
The amplitudes have been calculated in ref.\cite{papa:sngw}
using the E-vector formulation of the spinor-product
technique \cite{spinor}.
\begin{table}[htb]
\bc\begin{tabular}{|c|c|c|c|c|} \hline
$\sqrt{s}$ (GeV) & 170  & 190 & 230 & 500       \\ \hline
$We\nu$ (pb) & 1.1(1) & 1.4(1) & 1.3(1) & 0.55(5)\\ \hline
$WW$ (pb)    & 1.11(5) & 1.37(5) & 1.21(5) & 0.31(3)\\ \hline
\end{tabular}
\caption[.]{Cross sections in pb for different center-of-mass
energies.}
\label{tab2}
\ec\end{table}
The main advantage of this channel is that it does not
require the reconstruction of the produced $W$, which is,
in contrast, very important for the $W$-pair channel.
In order to exclude or minimize possible backgrounds
such as $b\bar{b}$ or $Z^0Z^0$ production it
is sufficient to
detect an electron well separated from the hadronic
activity (jets) and identify, using the overall
centre-of-mass energy, the existence of missing energy.
Still there is
the contribution of $e^+e^-\to e^{\pm}\bar{\nu}q\bar{q}'$ channel
when the invariant mass of the two quarks is well away from the
$W$ mass, but one expects that this is at least
an order of magnitude smaller than the
contribution of the single-$W$ channel.
\begin{table}[htb]
\bc\begin{tabular}{|c|c|c|c|c|c|c|} \hline
Bin number & 1 & 2 & 3 & 4 & 5 & 6      \\ \hline
$\sigma_{0,i}$& 0.87(2) & 1.22(3) & 1.76(4)
 & 2.65(7) & 4.2(1) & 8.1(2)   \\ \hline
$\sigma^*_{0,i}$& 0.87 & 1.22 & 1.73 & 2.57 & 4.20 & 8.31 \\ \hline
\end{tabular}
\caption[.]{Bin contents in pb; $\sigma^*_{0,i}$ are taken from
ref.\cite{bm200}.}
\label{tab3}
\ec\end{table}
\par In \tab{tab2}, I have
summarized the total cross-section rates for
$We\nu$ production, taking into account the following cuts
on the energy and the angle of the electron:
\bq E_e\ge 10\;\gev\;\;\; |\cos\vartheta_e|\le 0.95\;\;,\eq
as a function of the centre-of-mass
energy, from 170 to 500 GeV.
The branching ratio BR($W^\pm\to\mbox{hadrons})=2/3$ has been
taken into account.
The cross sections for $WW$ production are also displayed,
multiplied by the appropriate branching ratio ($2\over27$).
The calculation is done using the same method as for
the $We\nu$ channel \cite{papa:sngw}, and a cut on the $W$
angle, $|\cos~\vartheta_W|\le 0.98$ is considered.
Although this last cut is not necessary
for the calculation of the $WW$ cross section
(no singularities are present), it is included in order to
compare my results (see \tab{tab3})
with those of ref.\cite{bm200}, where
a detailed analysis of the $WW$ channel has been presented.
Assuming an integrated luminosity L=500~pb$^{-1}$, the events
expected from the $W^\pm e^\mp\nu$
channels are of the order of $\sim 1400$,
a rather good starting point for the angular-distribution analysis.
In order to quantify in a rather simple and transparent
way the deviation of the angular distribution from its
\SM prediction, I divide the interval, $-0.95\le\cos\vartheta\le0.95$,
in six equidistant bins and I calculate the $\chi^2$ as follows
\bq
\chi^2=\sum_{i=1,6} {(\sigma_{0,i}-\sigma_i)^2\over \delta\sigma_{0,i}^2}
\eq
where $\sigma_{0,i}$ is the \SM
cross section contained in the $i$th bin,
$\sigma_i$ is the corresponding cross section calculated
at non-zero deviation-parameter values
and $\delta\sigma_{0,i}$ represent the statistical errors
defined as
\bqa
\delta\sigma_{0,i}=\sqrt{\sigma_{0,i}\over\e \mbox{L}}
\nn\eqa
where $\e$ is the `efficiency' factor, which in this case is
simply 1, since no reconstruction of the $W$ is required.
For $WW$ production the analysis is the same, except that
the angle $\vartheta$ is taken to be the angle of $W^-$, being
reconstructed via its decay products. In order to compare the
two channels one has to consider an extra `efficiency' factor
taking into account
reconstruction uncertainties: I take as rather typical values
$\e=0.8$ for the total $WW$ cross section and $\e=0.64$ for
the angular distribution.
I also evaluate a more `theoretical' (and less
experimentally biased) $\chi^2_{th}$, defined as
\bqa
\chi^2_{th}={1\over 6}
\sum_{i=1,6} {(\sigma_{0,i}-\sigma_i)^2\over \sigma_{0,i}^2}
\label{c2th}
\nn\eqa
which is simply a measure of the relative deviation, in order
to have a comparison of the `theoretical' sensitivities on the
anomalous couplings.
\par
In \fig{fi1}, I show the angular distribution for single-$W$
as well as for $WW$ production. Both channels
exhibit similar
characteristics and they have, qualitatively,
the same sensitivity with respect to the anomalous couplings.
In \fig{fi2},
$\c2_{th}$ is presented as a function of the
deviation parameter $\xg$ (for the $\O_{W\Phi}$ operator).
As expected, angular distribution is more sensitive,
with respect to the deviation parameters,
than the total cross section. Also the angular distribution
of the $W$-boson angle ($\vartheta_W$)
is more sensitive than the corresponding quantity
for electrons. Of course in this `theoretical'
comparison, any problem concerning
$W$-reconstruction is simply ignored.
On the other hand, \fig{fi3} shows the $\c2$ as a function
of the same parameter, but now including all statistical
effects. It is possible to use the $\c2$ to estimate the region
in deviation-parameter-space, which is incompatible with
\SM predictions.
For the total cross section, limits are extracted
requiring that $\c2\ge 4$, which corresponds
to two standard deviations ($\sim95$\% CL). For the
angular distribution, limits are extracted requiring
that $\c2\ge 11.07$ at 95\% CL: for this the
corresponding number of degrees of freedom ($6-1=5$) is
considered \cite{eadi}.
It is worth while to emphasize that present results disagree with the
results of ref.\cite{bm200}, where, although not explicitly
written, an artificially small $\c2 \sim 1$ has been considered.
\begin{table}[htb]
\bc\begin{tabular}{|c|c|c|} \hline
Operator           & $We\nu$       & $WW$           \\ \hline
$\O_{W\Phi}$        &$-0.24$ $0.36$ & $-0.22$ $0.42$ \\ \cline{2-3}
$\xg$               &$-0.26$ $0.40$ & $-0.14$ $0.20$ \\ \hline
$\O_{B\Phi}$        &$-0.55$ $1.12$ & $-0.50$ $1.41$ \\ \cline{2-3}
$\xg$               &$-0.44$ $1.20$ & $-0.52$ $1.64$ \\ \hline
$\O_{W}$            &$-0.34$ $0.46$ & $-0.32$ $0.49$ \\ \cline{2-3}
$\lg$               &$-0.38$ $0.52$ & $-0.24$ $0.42$ \\ \hline
$\O^\prime_{W\Phi}$ &$-0.35$ $0.63$ & $-0.33$ $0.78$ \\ \cline{2-3}
$\xg$               &$-0.37$ $0.62$ & $-0.23$ $0.30$ \\ \hline
$\O^\prime_{W}$     &$-0.50$ $0.85$ & $-0.52$ $0.90$ \\ \cline{2-3}
$\lg$               &$-0.48$ $0.92$ & $-0.54$ $1.10$ \\ \hline
\end{tabular}
\caption[.]{Limits on deviation parameters. The first row
corresponds to the total cross section and the second row
to the angular-distribution analysis, as explained in the text.}
\label{tab4}
\ec\end{table}
Table \ref{tab4} summarizes the limits on the parameters describing
the deviations from the \SM predictions
both for $WW$ and single-$W$ production.
Although the results
from the angular distribution of $W$ are in general better,
the gain cannot be considered significant, since they
cover a rather small and less important region in the parameter space.
Furthermore, if reconstruction efficiencies for $W$ are lowered
by a factor $\sim0.8$, the above-mentioned region disappears.
On the other hand, if reconstruction uncertainties
are overcome, the limits are
not essentially affected: if instead
of $\e=0.64$ one takes $\e=1$, the limits for the $\O_{W\Phi}$
operator, for instance, change only from $-0.14\le \xg\le 0.20$ to
$-0.12\le\xg\le 0.16$.
\par In summary, I have shown that single-$W$ production
is essentially as sensitive as the $WW$ channel with respect
to the trilinear couplings of vector bosons. The reconstruction
of the $W$ will not play an important role in limiting
the anomalous couplings, at least for one-parameter CP-conserving
deviations.
Single-$W$ production at $\sqrt{s}=190\;\gev$
will be able to set limits of the order of $0.2 - 1$ on the
anomalous couplings, improving by a factor of $5 - 10$ the
existing limits \cite{anom:ua2} and by a factor of $2 - 3$ those
expected from HERA \cite{hera}.

\begin{figure}[p]
\begin{center}
\mbox{\psfig{file=sw_an.eps.a,width=12cm,height=12cm}}
\caption[.]{Angular distribution for $We\nu$ and $WW$ channels.
The `experimental' points correspond to the \SM predictions
and the errors are
purely statistical. The dot-dashed lines correspond to $\xg=1$
and the dashed ones to $\xg=0.5$.}
\label{fi1}
\end{center}
\end{figure}

\begin{figure}[p]
\begin{center}
\mbox{\psfig{file=sw_xth.eps.a,width=7cm,height=7cm}}
\caption[.]{The $\c2_{th}$ function, defined in \eqn{c2th},
for the operator $\O_{W\Phi}$.
For the $We\nu$ ($WW$) channel,
the solid (dot-dashed) line corresponds to the total cross section
and the dashed (dotted) line to the angular distribution.}
\label{fi2}
\end{center}
\end{figure}

\begin{figure}[p]
\begin{center}
\mbox{\psfig{file=sw_x.eps.a,width=7cm,height=7cm}}
\caption[.]{The $\c2$ function for the operator $\O_{W\Phi}$.
For the $We\nu$ ($WW$) channel,
the solid (dot-dashed) line corresponds to the total cross section
and the dashed (dotted) line to the angular distribution.}
\label{fi3}
\end{center}
\end{figure}

\end{document}